\documentclass{ws-procs975x65}

\usepackage{amssymb}
\usepackage{amsmath}
\usepackage[usenames,dvipsnames]{color}
\usepackage{hyperref}
\usepackage{graphicx}
\usepackage[dvips]{epsfig}
\usepackage[usenames,dvipsnames]{color}

\hypersetup{
    colorlinks=true,         % false: boxed links; true: colored links, false is default
    linkcolor=blue,          % color of internal links, red is default
    citecolor=red,        % color of links to bibliography, 'green' is default
    urlcolor=Violet             % color of external links, cyan is default
}

\def\be{\begin{equation}}
\def\ee{\end{equation}}

\begin{document}

\title{\uppercase{Differences and similarities between Shape Dynamics and General Relativity}}
\author{\uppercase{Henrique Gomes$^1$, Tim A. Koslowski$^2$}}
\address{$^1$ University of California at Davis, One Shields Avenue Davis, CA, 95616, USA\\$^2$ University of New Brunswick, Fredericton, NB, E3B 5A3, Canada\\ email: $^1$\texttt{gomes.ha@gmail.com}, $^2$\texttt{t.a.koslowski@gmail.com}}

\bodymatter

\begin{abstract}
The purpose of this contribution is to elucidate some of the properties of Shape Dynamics (SD) and is largely based on the longer article \cite{FAQ}. We shall point out some of the key differences between SD and related theoretical constructions, illustrate the central mechanism of symmetry trading in electromagnetism and finally point out some new quantization strategies inspired by SD. We refrain from describing mathematical detail and from citing literature. For both we refer to \cite{FAQ}.  
\end{abstract}

{\bf What is Shape Dynamics?}
SD is a Hamiltonian description of General Relativity (GR) on the familiar ADM phase space that is locally indistinguishable form ADM formulation, but based on a different gauge invariance: The local refoliation invariance of the ADM description is replaced by local spatial Weyl invariance in SD while both descriptions posses spatial diffeomorphism invariance. This ``symmetry trading'' is due to the redundancy of degrees of freedom in gauge theories and can be straightforwardly constructed using a ``linking gauge theory.''\\
A linking theory is a theory that contains an auxiliary canonical pair $(\phi_\alpha,\pi^\alpha)$ and a larger gauge symmetry and admits $\phi_\alpha\equiv 0$ and $\pi^\alpha\equiv 0$ as two distinguished gauge fixings. The phase space reduction w.r.t. either set of conditions eliminates the auxiliary canonical pair and the Dirac bracket associated with the phase space reduction coincides with the Poisson bracket on the original phase space. It is clear that the two partially gauge fixed theories describe the same physics. This can be explicitly seen through a further gauge fixing of the two theories, because it turns out that one can always find gauge fixings  such that the initial value problem and the equations of motion of both theories coincide; this completely gauge fixed theory acts as a ``dictionary'' between the two descriptions. An alternative construction of equivalent gauge theories starts with the dictionary followed by two partial gauge--unfixings.\\
In the case of GR, this alternative route starts with the constant mean extrinsic curvature (CMC) gauge of the ADM description, which can be unfixed in two ways: (1) to the standard ADM description and (2) to the SD description of GR. The local availability of CMC gauge in GR explains why ADM and SD are locally indistinguishable. Globally however, one finds two obstructions: on the one hand, a GR solution may not admit a global CMC foliation, and on the other hand, the translation of a perfectly well defined SD solution into the spacetime description may lead to a degenerate spacetime geometry.\\
{\bf Construction and content of Shape Dynamics:} 
The dictionary between the ADM and SD formulation of GR is ADM in CMC gauge, which plays an important role in the initial value problem\cite{York}. Using the spatial metric $g_{ab}$ and ADM momenta $\pi^{ab}$ on a compact Cauchy surface $\Sigma$ without boundary, one can write the CMC gauge condition as $\pi(x)-\langle \pi \rangle\sqrt{|g|(x)}=0$, where $\pi=g_{ab}\pi^{ab}$ and $\langle \pi \rangle=\int \pi/\int \sqrt{|g|}$. This gauge condition gauge-fixes all local Hamiltonian constraints of the ADM formulation and leaves only one global volume constraint $\int d^3x \sqrt{|g|}(1-\Omega_o^6[g,\pi])\approx 0$ unfixed, where $\Omega_o[g,\pi]$ denotes the solution to the Lichnerowicz-York equation. SD is obtained as a gauge-unfixing of the CMC condition, which turns out to be the generator of local Weyl--transformations that preserve the total spatial volume. The variable $\tau=\frac 3 2 \langle \pi\rangle$ is identified with York time and $V$ with its conjugate momentum. After deparametrizing this time variable,
 one obtains first class constraints
\begin{equation}
 D(\rho)=\int_\Sigma d^3x \rho(\pi-\frac 2 3 \tau \sqrt{|g|}),\,\,\,\,H(v)=\int_\Sigma d^3x \pi^{ab}(\mathcal L_v g)_{ab},
\end{equation}
which generate spatial Weyl-transformations and spatial diffeomorphisms. The physical Hamiltonian turns out to be the conformal York Hamiltonian:
\begin{equation}
 H_{SD}=\int_\Sigma d^3x\sqrt{|g|}\, \Omega_o^6[g,\pi;\tau).
\end{equation}
{\bf Example of symmetry trading: shift symmetry in Electromagnetism.}
The symmetry trading mechanism used to construct SD can be applied to general gauge theories. One of the simplest examples is pure electromagnetism. For this we assume boundary conditions for standard electromagnetism that ensure that axial gauge $A_3(x)\equiv 0$ is a complete gauge-fixing of the Gauss constraints. In this case one can construct a linking theory as
\begin{equation}
 \begin{array}{c}
   H=\int d^3x \,\frac 1 2 \left( \delta_{ab}E^aE^b+\delta^{ab}B_aB_b+(\phi^2+2E^3\phi) \right)\\
   G(\Lambda)=\int d^3x \,\Lambda \left(E^a_{,a}+\phi_{,3}\right)\approx 0,\,\,\,\, Q(\rho)=\int d^3x \,\rho \left(\pi_\phi-A_3\right)\approx 0,
 \end{array}
\end{equation}
where $A_a,E^a$ denote the vector potential and the electric field, while $\phi,\pi_\phi$ denote an auxiliary pair. The special gauge $\phi(x)\equiv 0$ leads to the phase space reduction $(\phi,\pi_\phi)\to(0,A_3)$. This trivializes the constraints $Q(\rho)$ and turns the constraints $G(\Lambda)$ and the Hamiltonian $H$ into the Gauss constraint and the Hamiltonian of electrodynamics.\\
The gauge condition $\pi_\phi(x)\equiv 0$ leads to the phase space reduction $(\phi,\pi_\phi)\to(F[E^1,E^2]-E^3,0)$, where $F[E^1,E^2]=\int^{x_3} dt(E^1_{,1}(x_1,x_2,t)+E^2_{,2}(x_1,x_2,t))$. The resulting shift symmetric theory is described by the shift constraints and Hamiltonian
\begin{equation}
 \begin{array}{rcl}
   Q(\rho)&=&\int d^3x\,\rho\,A_3,\\
   H_{shift}&=&\int d^3x\,\frac 1 2 \left((E^1)^2+(E^2)^2+(\vec B)^2+F[E_1,E^2]^2\right).
 \end{array}
\end{equation}
This simple example exhibits  two analogies with the symmetry trading between ADM and SD: (1) A simple Hamiltonian and a complicated set of constraints is traded for a simple set of constraints and a complicated Hamiltonian. This feature is negligible in electrodynamics, because one knows how to treat Gauss constraints effectively. The technical advantage can thus not be compared with with ADM Hamiltonian constraints, which are nonlinear in $g_{ab}$ and $\pi^{ab}$ and moreover depend on second derivatives of $g_{ab}$. Symmetry trading between ADM and SD is thus a much more valuable technical simplification.\\
(2) The dual theory has an ultralocal initial value problem and a local Poisson bracket, while the dictionary theory (electromagnetism in axial gauge) posses a nonlocal Dirac bracket and an initial value problem that requires the solution of a differential equation. These features are completely analogous to the relation of SD and ADM in CMC gauge.\\
{\bf Some answers to frequently asked questions about SD:}\\
%\begin{enumerate}
 1. {\it Dynamical Equivalence:} ADM and SD possess identical Poisson algebras of observables and identical evolution equations for these observables as long as CMC gauge is available. However, as we explained in the opening section, the two theories can differ if CMC gauge breaks down in the ADM description or if the reconstructed spacetime geometry that corresponds to an SD solution is degenerate. This is the source of real, observable physical differences between ADM and SD.\\
 2. {\it Difference with gauge-fixed ADM:} SD is a gauge theory of spatial diffeomorphisms and Weyl transformations on unreduced ADM phase space, which distinguishes it form the ADM formulation and from any gauge fixing of the ADM formulation.\\  
 3. {\it Relation with AdS/CFT:} The bulk/bulk equivalence of Shape Dynamics and ADM reduces in a large volume limit to terms known from holographic renormalization. \\
 4. The initial value problem for Shape Dynamics is significantly simpler than the initial value problem for ADM; it is solved by finding transverse traceless momenta.\\
 5. We explicitly recover Poincar\'e invariance of a set of Shape Dynamics data that represents Minkowski space.  \\
 6. Straightforward attempts to quantize Shape Dynamics as wave functions of the metric are very similar to the analogous attempts to quantize ADM in CMC gauge, but with the important difference that one works with an unreduced phase space. This means in particular that one can impose canonical commutation relations coming from the Poisson bracket rather than non-canonical commutation relations coming form a Dirac bracket.  \\
 7. From an effective field theory point of view, where field content and symmetries of the action are the only ingredients, Shape Dynamics represents a radical departure from GR. Based on SD, one looks for spatial diffeomorphism- and Weyl- invariant operators, as opposed to space-time diffeomorphism invariant ones. 
%\end{enumerate}
\\
{\bf Acknowledgments:}\\
 HG was supported in part by the U.S.
Department of Energy under grant DE-FG02-91ER40674. TK was supported by NSERC.

\end{document}